\documentclass[aps,preprint,nofootinbib,floatfix]{revtex4-1}
\usepackage{graphicx,color,hyperref}
\usepackage{amsmath,amssymb}
\usepackage{url}
\usepackage{epstopdf}
\newcommand{\lsim}{\mathrel{\mathop{\kern 0pt \rlap
  {\raise.2ex\hbox{$<$}}}
  \lower.9ex\hbox{\kern-.190em $\sim$}}}
\newcommand{\gsim}{\mathrel{\mathop{\kern 0pt \rlap
  {\raise.2ex\hbox{$>$}}}
  \lower.9ex\hbox{\kern-.190em $\sim$}}}

\begin{document}

\title{Collider Signatures of Goldstone Bosons}

\author{Kingman Cheung$^{1,2}$, Wai-Yee Keung$^{3}$ and Tzu-Chiang Yuan$^4$}

\affiliation{
$^1$Division of Quantum Phases \& Devices, School of Physics, 
Konkuk University, Seoul 143-701, Republic of Korea \\
$^2$Department of Physics, National Tsing Hua University, 
Hsinchu 300, Taiwan\\
$^3$Physics Department, University of Illinois at Chicago, 
Chicago IL 60607, U.S.A \\
$^4$Institute of Physics, Academia Sinica, Nankang, Taipei 11529, Taiwan
}

\date{\today}

\begin{abstract}
Recently Weinberg suggested that Goldstone bosons arising from the
spontaneous breakdown of some global hidden symmetries can interact
weakly in the early Universe and account for a fraction of the
effective number of neutrino species $N_{eff}$, which has been
reported persistently 1$\sigma$ away from its expected value of three.
In this work, we study in some details a number of experimental
constraints on this interesting idea based on the simplest possibility
of a global $U(1)$, as studied by Weinberg.  We work out the decay
branching ratios of the associated light scalar field $\sigma$ and
suggest a possible collider signature at the Large Hadron Collider
(LHC).  In some corners of the parameter space, the scalar field
$\sigma$ can decay into a pair of pions with a branching ratio of
order $O(1)\%$ while the rest is mostly a pair of Goldstone bosons.  The
collider signature would be gluon fusion into the standard model Higgs 
boson $gg \to H$ or associated production with a $W$ gauge boson 
$q \bar q' \to H W$, followed by $H \to \sigma \sigma \to (\pi\pi) (\alpha\alpha)$
where $\alpha$ is the Goldstone boson.
\end{abstract}

\maketitle


\section{Introduction}
Existence of Goldstone boson(s) is a manifestation of the
spontaneous breakdown of some exact or nearly-exact global continuous
symmetry in Nature \cite{goldstone}.  Such Goldstone or pseudo-Goldstone bosons 
would be exactly or nearly massless. 
The well known example in the standard model
(SM) is the pion which in the modern view can be interpreted as 
the Goldstone boson of spontaneous breakdown of the chiral 
$SU(2) \times SU(2)$ symmetry.
Another logical possibility is the presence of a global
hidden symmetry that the usual SM particles do not
experience. The simplest choice is a global hidden $U(1)$ 
symmetry associated with a new quantum number $W$ of which all
the hidden particles carry non-vanishing $W$ charges
while all SM particles are neutral.

Weinberg \cite{weinberg} showed that such a simple extension to the SM
could bring the Goldstone boson into weak interactions with the SM
particles via a Higgs portal, $g (S^\dagger S)(\Phi^\dagger \Phi)$, 
where $S(x)$ is a complex singlet scalar field neutral under the SM
symmetries with a nonzero $W$ quantum number, and $\Phi$ is the SM Higgs 
doublet with $W=0$. 
Thus, the Goldstone bosons could remain in thermal equilibrium in the early 
Universe until they went out of equilibrium at a 
temperature above but not much above the muon mass.
In this way, the Goldstone boson could contribute a fraction of
$0.39$ to the effective number $N_{eff}$ of neutrino species present
in the era before recombination \cite{weinberg}.
The requirement for this to happen is that the interactions of 
the Goldstone boson with the SM particles should be strong enough to bring
it into thermal equilibrium and also weak enough such that it decouples
close to the neutrino-decoupling temperature. The nature of derivative 
couplings of
Goldstone bosons can easily satisfy this requirement.

There are a number of constraints on the model, namely on the Goldstone
boson and the massive scalar  $\sigma$ field associated with the Goldstone
boson.  Since they are weakly coupled to the Higgs boson, they would 
contribute to the invisible decay width of the Higgs boson 
\cite{weinberg,Huang:2013oua}. There are a number of other constraints
in existing data \cite{Huang:2013oua}, e.g., search for invisible 
particles in hadron decays, quarkonium decays, etc.
In particular, here we point out that
the invisible Higgs search at LEP-II will give the most stringent
constraint on the mixing angle. The detail will be given in the next section.

In this work, besides working out the constraints on the model, 
we point out it may be possible to detect the $\sigma$ field and
the Goldstone boson of the model at the LHC, via the {\it visible} decay
mode of the $\sigma$ field, namely $\sigma \to \pi \pi$, 
especially when the modulus of the field $S$ takes on a large
vacuum expectation value (VEV). This is the main result of this work.
We also estimate the event rates at the LHC-8 and LHC-14. 

Studies of Goldstone bosons at the LHC in other context can be found 
in Ref.~\cite{dedes}. The related dark matter phenomenology has also been
studied in Ref.~\cite{luis}.

\section{The Model}
The model \cite{weinberg} is based on adding a
complex singlet field $S$ to the SM Higgs doublet, through which the
singlet field interacts with the SM particles. The renormalizable
Lagrangian density is given by 
\footnote{
We have normalized the kinetic energy term of a complex scalar field
in the canonical form with the coefficient equals to 1.}
\begin{equation}
\label{1}
{\cal L}= (\partial_\mu S^\dagger ) (\partial^\mu S ) + \mu^2 S^\dagger S 
- \lambda (S^\dagger S)^2  - g (S^\dagger S) (\Phi^\dagger \Phi) 
+ {\cal L}_{\rm sm}
\end{equation}
where the Higgs sector in the ${\cal L}_{\rm sm}$ is
\begin{equation}
{\cal L}_{\rm sm} \supset (D_\mu  \Phi)^\dagger ( D^\mu \Phi) + \mu^2_{\rm sm}
 \Phi^\dagger \Phi - \lambda_{\rm sm} (\Phi^\dagger \Phi )^2 \; .
\end{equation}
To respect the low energy theorem, we follow Ref.~\cite{weinberg} to 
write $S$ in term of a radial field $r(x)$ and a Goldstone field 
$\alpha(x)$ as 
\begin{equation}
 S(x) = \frac{1}{\sqrt{2}} \left( \langle r \rangle + r(x) \right) 
e^{i 2 \alpha(x)}
\end{equation}
in which the radial field develops a VEV
$\langle r \rangle$
where the field $S$ is expanding around. Note that one can always set 
$\langle \alpha(x) \rangle =0$
by field redefinition.
The SM Higgs doublet field $\Phi$ is expanded about the VEV as
\begin{equation}
 \Phi (x) = \frac{1}{\sqrt{2}}\, \left( \begin{array}{c}
                           0 \\
                         \langle \phi \rangle + \phi (x)
                                      \end{array}
  \right )
\end{equation}
in the unitary gauge, and $\langle \phi \rangle \approx 246$ GeV.
Expanding the Lagrangian in Eq.~(\ref{1}) around the VEVs and replacing
$\alpha(x) \to \alpha(x)/(2\langle r \rangle )$ in order to achieve a canonical
kinetic term of the $\alpha(x)$ field describing a scalar of dimension 1, 
we obtain
\begin{eqnarray}
\label{L_W}
{\cal L} & \supset & \frac{1}{2} (\partial_\mu r )(\partial^\mu r) 
  + \frac{1}{2} \frac{ (\langle r \rangle + r)^2}{\langle r \rangle^2}
   (\partial_\mu \alpha )(\partial^\mu \alpha) 
 + \frac{\mu^2}{2} ( \langle r \rangle + r )^2 
 - \frac{\lambda}{4} ( \langle r \rangle + r )^4 \nonumber \\
&&+ \frac{1}{2} (\partial_\mu \phi )(\partial^\mu \phi)
 + \frac{\mu_{\rm sm}^2}{2} ( \langle \phi \rangle + \phi )^2 
 - \frac{\lambda_{\rm sm}}{4} ( \langle \phi \rangle + \phi )^4 \nonumber \\
&& - \frac{g}{4} ( \langle r \rangle + r )^2 ( \langle \phi \rangle + \phi )^2 \; .
\end{eqnarray}
Two tadpole conditions can be written down using $\partial V / \partial r=0$
and $\partial V / \partial \phi=0$ where $V$ is the scalar potential part 
of Eq.(\ref{L_W}):
\begin{eqnarray}
\langle \phi \rangle^2 &=& 
 \frac{4 \lambda\mu_{\rm sm}^2 - 2 g \mu^2 }
   { 4 \lambda \lambda_{\rm sm} - g^2 } \;, \\
\langle r \rangle^2 &=& \frac{4 \lambda_{\rm sm}\mu^2 -2 g \mu_{\rm sm}^2 }
 { 4 \lambda \lambda_{\rm sm} - g^2 }  \; .
\end{eqnarray}
Taking the decoupling limit $g\to 0$ from the above equations, we
recover the SM condition of 
$\langle \phi \rangle^2 = \mu_{\rm sm}^2/\lambda_{\rm sm}$
as well as $\langle r \rangle^2 = \mu^2/\lambda$.

The interaction fields $r(x)$ and $\phi(x)$ are no longer mass eigenstates
because of the mixing term proportional to $g$. The mass term is
\begin{equation}
{\cal L}_{\rm m} = - \frac{1}{2}  \left( \phi(x) \;\;\; r(x) \right)\;
  \left ( \begin{array}{cc} 
  2 \lambda_{\rm sm} \langle \phi \rangle^2 
             & g \langle r \rangle \langle \phi \rangle\\
               g \langle r \rangle \langle \phi \rangle &
  2 \lambda \langle r \rangle^2 \end{array} \right )\;
  \left( \begin{array}{c} 
            \phi(x) \\
             r(x) \end{array}  \right ) \;.
\end{equation}
We rotate $(\phi(x)\;\; r(x) )^T$ by an angle $\theta$ into physical fields:
\begin{equation}
 \left( \begin{array}{c}
             H(x) \\
             \sigma(x) \end{array} \right ) = 
 \left( \begin{array}{cc}
         \cos\theta & \sin\theta \\
         - \sin\theta & \cos\theta  \end{array} \right )\;
 \left( \begin{array}{c}
             \phi(x) \\
              r(x) \end{array} \right ) \;.
\end{equation}
The physical masses of the $H(x)$ and $\sigma(x)$, and the mixing angle
are given by
\begin{eqnarray}
 m_H^2 &=& 2 \lambda_{\rm sm} \langle \phi \rangle^2 \cos^2\theta 
         + 2 \lambda  \langle r  \rangle^2 \sin^2\theta 
         + g  \langle r  \rangle  \langle \phi  \rangle \sin 2 \theta \; ,
  \nonumber \\
 m_\sigma^2 &=& 2 \lambda \langle r \rangle^2 \cos^2\theta 
         + 2 \lambda_{\rm sm}  \langle \phi  \rangle^2 \sin^2\theta 
         - g  \langle r  \rangle  \langle \phi  \rangle \sin 2 \theta \; ,
 \\
\tan 2 \theta &=& \frac{ g  \langle r  \rangle  \langle \phi  \rangle}
   {\lambda_{\rm sm}  \langle \phi  \rangle^2 - \lambda \langle r  \rangle^2} \; .
   \nonumber
\end{eqnarray}
In the small $\theta$ limit ($\theta \alt 0.01$ as will be shown later),
\begin{eqnarray}
\label{smalltheta}
m_H^2 &\approx & 2 \lambda_{\rm sm} \langle \phi \rangle^2 \; , \nonumber \\
m_\sigma^2 &\approx& 2 \lambda \langle r \rangle^2 \; ,  \\
\theta &\approx & \frac{ g  \langle r  \rangle  \langle \phi  \rangle}
  { m_H^2 - m_\sigma^2 } \; . \nonumber
\end{eqnarray}
We can now write down the interactions terms in the limits of 
$\theta \ll 1$ and $m_\sigma \ll m_H$:
\footnote{
In the coupling of $H\sigma\sigma$, the next-to-leading term in $\theta$ is 
$g \theta \langle r \rangle$, which is suppressed by a factor of  
$(\langle r \rangle / \langle \phi \rangle) \theta $ relative to the leading
term. However, when the ratio $\langle r \rangle/ \langle \phi \rangle $ is 
large, this next-to-leading term could be a sizable correction to the 
leading term.
}
\begin{eqnarray}
{\cal L}_{H \alpha\alpha} &=& \frac{\theta}{\langle r \rangle } \, H\,
  (\partial_\mu \alpha)   (\partial^\mu \alpha)  \; , \nonumber \\
{\cal L}_{\sigma \alpha\alpha} &=& \frac{1}{\langle r \rangle } \, 
 \sigma  \,
  (\partial_\mu \alpha)   (\partial^\mu \alpha)  \; , \\
{\cal L}_{H\sigma\sigma} &=&  - \frac{g}{2} \langle \phi \rangle \, H \,\sigma^2 \; . \nonumber
\end{eqnarray}

\section{Current Constraints on the $\sigma$ field and Goldstone
Boson}

{\it Constraints from LEP searches for an invisibly decaying Higgs boson.} 
The $\sigma$ field has a 
suggested mass of about 500 MeV and it can be produced via the 
mixing with the Higgs field \cite{weinberg}.
Such a light scalar boson can be
readily produced in hadron decays, quarkonium decays, as well as in the 
$Z$ boson decays, and at $e^+ e^-$ collisions. The OPAL collaboration 
\cite{opal}
searched for an invisibly decaying Higgs boson for the whole mass 
range from 1 GeV to 108 GeV at LEPII where the limit of the following ratio
\[
 \frac{ \sigma(Z h) B(h \to \chi^0 \chi^0)}{ \sigma (Z H_{\rm sm})} 
\]
was obtained.
We can extrapolate the mass of the $h$ boson to be below 1 GeV, and the 
ratio being excluded is down to almost $10^{-4}$ (See Fig.~5 of \cite{opal}). 

In the present context, the production cross section of $Z \sigma$ is
\[
 \sigma (Z \sigma ) = \theta^2 \times \sigma (Z H_{\rm sm} ) 
\]
Assuming the $\sigma$ field decays entirely into Goldstone bosons and
disappears, we can constrain the mixing angle $\theta$
\begin{equation}
 \theta \alt 10^{-2} \;.
\end{equation}
A less stringent constraint $\theta<0.27$ has also been obtained recently in
\cite{Huang:2013oua}
from $B(\Upsilon \to\gamma  E) \hskip-12pt / \; < 2\times 10^{-6}$
\cite{delAmoSanchez:2010ac}
via Wilczek mechanism \cite{Wilczek:1977zn}
with the  one-loop QCD correction \cite{Nason:1986tr}.

{\it Invisible width of the Higgs boson}. The invisible decay of the
Higgs boson goes through two processes:
\[
 H \to \alpha \alpha,\qquad 
 H \to \sigma \sigma \to 4 \alpha
\]
The decay partial widths are given by
\begin{eqnarray}
\Gamma(H \to \alpha\alpha) &=& \frac{1}{32\pi} 
    \frac{m_H^3}{ \langle  \phi \rangle^2 } \,
   \frac{  \langle  \phi \rangle^2}{\langle  r \rangle^2 } \, \theta^2
 \; , \\
\Gamma(H \to \sigma\sigma) & \approx & \frac{1}{32\pi} 
    \frac{m_H^3}{ \langle  \phi \rangle^2 } \,
   \frac{  \langle  \phi \rangle^2}{\langle  r \rangle^2 } \, \theta^2 \; ,
\end{eqnarray}
in which we have assumed $m_\sigma \ll m_H$.  
Since the $\sigma$ field decays mostly into the Goldstone bosons (see
the next section), we add both channels to obtain the invisible width
of the Higgs boson \cite{wimpmode}
\begin{equation}
\label{Hinv}
\Gamma_{\rm inv} (H) = \frac{2}{32\pi} 
    \frac{m_H^3}{ \langle  \phi \rangle^2 } \,
   \frac{  \langle  \phi \rangle^2}{\langle  r \rangle^2 } \, \theta^2 \; .
\end{equation}

\begin{figure}[th!]
\centering
\includegraphics[width=5.5in]{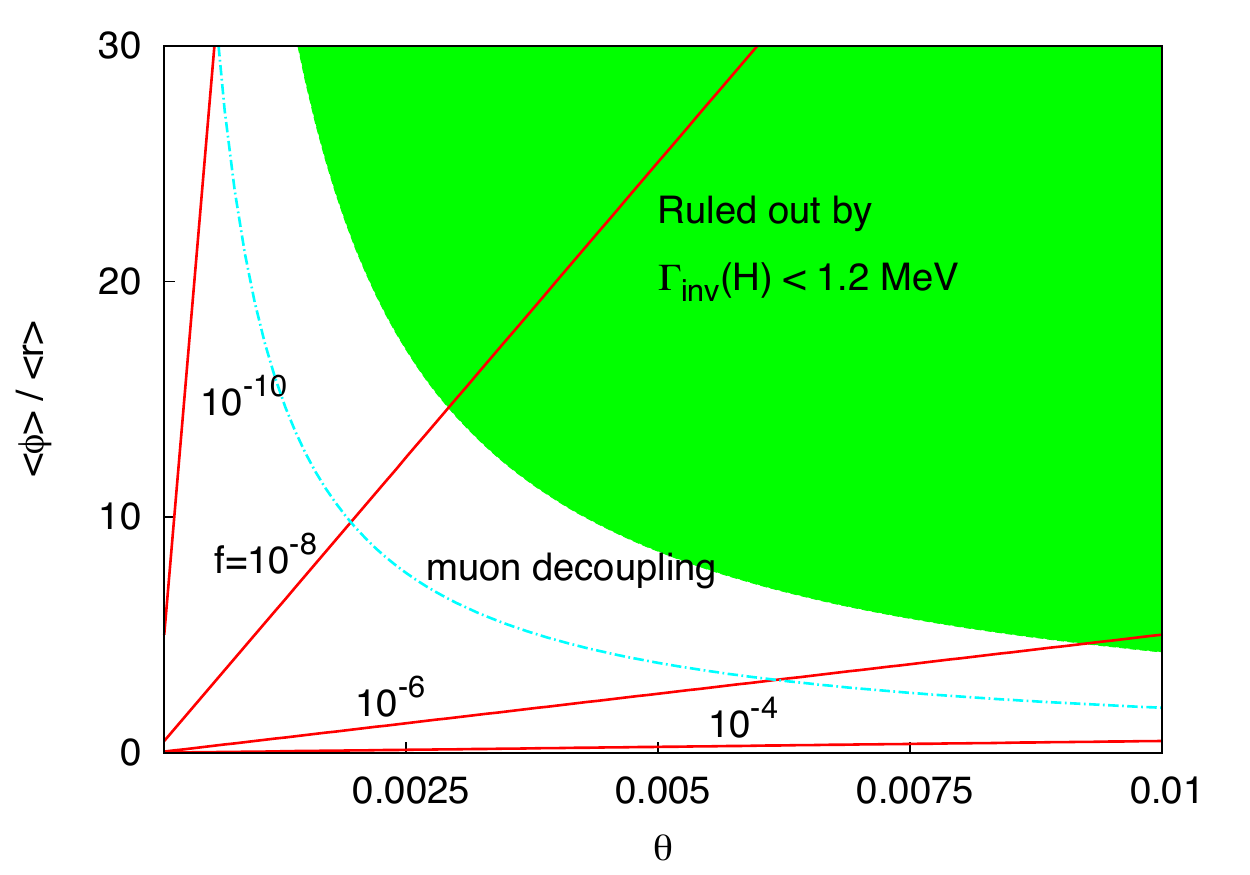}
\caption{\small \label{limit}
Parameter space of $(\theta, \langle \phi \rangle /\langle r \rangle)$.
The shaded region is ruled out by the invisible width of the Higgs boson to
be less than 1.2 MeV. 
The condition for muon decoupling and the ratio $f$ 
are given in Eqs.~(\ref{muondecoupling}) and (\ref{f}),  respectively,
and are shown here for $m_\sigma = 500$ MeV.
The upper limit of $\theta$ is taken to be $0.01$ 
constrained by the search for invisibly decaying Higgs boson at LEP-II.
}
\end{figure}

One of the global fits to all the SM Higgs boson signal strength has constrained
the non-standard decay width of the Higgs boson to be less than $1.2$ MeV
(branching ratio about 22\%) at 95\% CL \cite{higgcision}.
The constraint on $\langle \phi\rangle / \langle r \rangle$ from the 
invisible width is similar to the constraint on $g$ obtained in 
Ref.~\cite{weinberg}.
Note that we also account for the decay of $H\to \sigma\sigma$ as a part of
the invisible width of the Higgs boson.
Numerically, from Eq.(\ref{Hinv}), we have
 \[
  \theta \; \frac{\langle \phi \rangle}{\langle r \rangle } \le 0.043 \;.
\]
We use this constraint to rule out the parameter space in the plane of
$(\theta, \langle \phi \rangle /\langle r \rangle)$, shown by the shaded
region in Fig.~\ref{limit}.

{\it Muon decoupling.}  In the early Universe, the
  Goldstone bosons are at thermal equilibrium with the SM
  particles. As the Universe cooled down from its Hubble expansion,
  the Goldstone bosons would go out of the equilibrium since its weak
  interaction with the SM particles could no longer keep up with the
  Hubble expansion. It was argued in \cite{weinberg} that the best
  scenario for the Goldstone bosons to go out of equilibrium is at a
  temperature still above the muon and electron masses but below all
  other masses of the SM. After decoupling, the Goldstone bosons were free and 
  its temperature $T$ would then just fall off like the inverse of the
  Friedmann-Roberston-Walker scale factor $a$. Since the total cosmic entropy
  is conserved during the adiabatic expansion, after the muon
  annihilation, the constancy of $Ta$ for the Goldstone bosons implies
  they behave like neutrino impostors contributing to the measured
  $\Delta N_{eff} = (4/7)(43/57)^{4/3}=0.39$ \cite{weinberg}, 
  which is consistent with the 
  recent Planck result \cite{planck}. For this
  scenario to work, the annihilation rate of $\alpha
  \alpha \longleftrightarrow \mu^+ \mu^-$ must be of the same order of the Hubble expansion
  rate at the temperature $k_B T \approx m_{\mu}$, {\it i.e.}
  \cite{weinberg}
\begin{equation}
\label{muondecoupling}
\frac{g^2 m_{\mu}^7 m_{\rm PL}}{m_{\sigma }^4 m_{H}^4} \approx 1 \; ,
\end{equation}
where $m_{\rm PL}$ is the Planck mass.
{}From Eq.(\ref{smalltheta}), one can express $g^2$ in terms of $\theta$, 
$\langle \phi \rangle / \langle r \rangle$ and $m_\sigma$. 
However, its dependence on $m_\sigma$ is rather weak for $m_\phi \gg m_\sigma$. 
This muon decoupling 
condition of Eq.(\ref{muondecoupling}) is shown in 
Fig.~\ref{limit}
for $m_\sigma = 500$ MeV. 
Note that this muon decoupling is not a constraint, 
but rather an interesting condition at the early universe for
the Goldstone boson to explain $\Delta N_{eff}$.

\section{Decay of the $\sigma$ field}

Because of the constraint from the Higgs invisible width and 
condition for muon decoupling in Eq.~(\ref{muondecoupling}), 
the mass range of $\sigma$ 
cannot be much larger than $O(1)$ GeV \cite{weinberg}. 
We therefore show the mass range
from 1 MeV to 1000 MeV for $\sigma$ from now on, and use 500 MeV when we 
need a typical value.
The decay modes of such a light $\sigma$ field are very similar to those
of a very light Higgs boson ($\alt 1$ GeV) \cite{pipi}. The $\sigma$ can 
decay into a pair of electrons, muons, photons, pions and Goldstone bosons.

The formulas for the decays into $e^+ e^-$, $\mu^+ \mu^-$ and $\gamma\gamma$
are the same as the Higgs boson, up to a mixing angle. 
Thus, for the $f \bar f$ final state, we have
\begin{equation}
\Gamma(\sigma \to f\bar f) = \theta^2 \; \frac{m_f^2 m_\sigma}{8 \pi \langle
\phi \rangle^2 }\, \left [ 1- \frac{4 m_f^2}{m_\sigma^2} \right ]^{3/2} \;,
\end{equation}
in the small $\theta$ limit. For $m_\sigma < 1$ GeV the only possibility for
fermionic decays are $f = e,\mu$. 
The decay width for $\sigma \to \gamma\gamma$ is the same as the one for the
SM Higgs boson, up to $\theta^2$. 
We do not repeat the formula here except noting
that the loop formulas for the light quarks are not trustworthy due to 
non-perturbative effects.
However the $\sigma \to \gamma\gamma$ mode is not important.

Since the $\sigma$ field is very light and close to the hadronic scale 
$\Lambda_{\rm QCD}$,
its decay into two gluons 
is not applicable because of the non-perturbative hadronic effects, 
in contrast to the SM Higgs boson. The only hadrons that the $\sigma$ 
field can decay into is a pair of pions $\pi^+ \pi^-$ and $\pi^0 \pi^0$.
The decay width of $\sigma \to \pi \pi$ summing
over the isospin channels is given by \cite{pipi}
\begin{equation}
\Gamma ( \sigma \to \pi \pi) = \theta^2\; \frac{1}{216 \pi}\,
   \frac{m_\sigma^3}{\langle \phi \rangle^2}\,
  \left( 1- \frac{4 m_\pi^2}{m_\sigma^2} \right )^{1/2} \,
  \left( 1 +  \frac{11 m_\pi^2}{2 m_\sigma^2} \right )^{2}  \;.
\end{equation}

\begin{figure}[th!]
\centering
\includegraphics[width=3.2in]{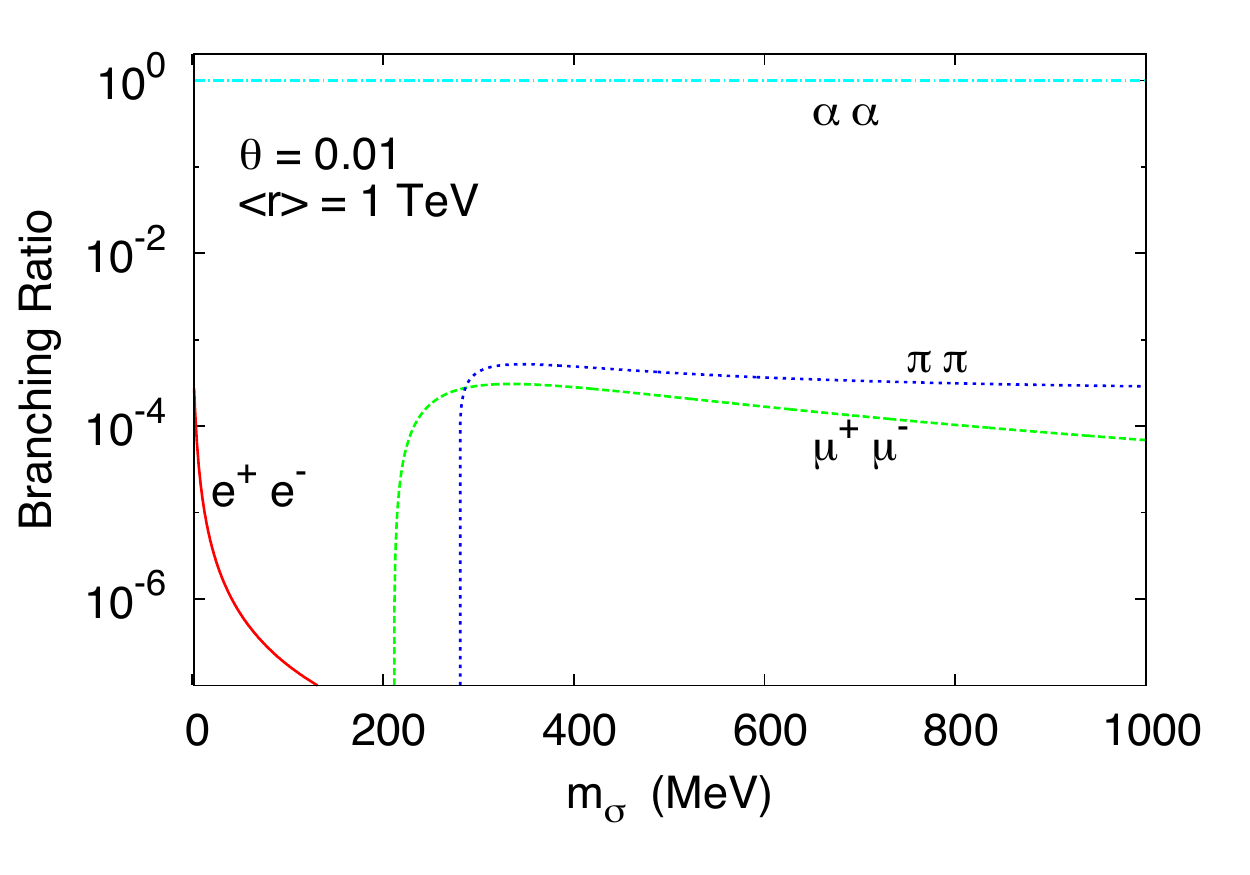}
\includegraphics[width=3.2in]{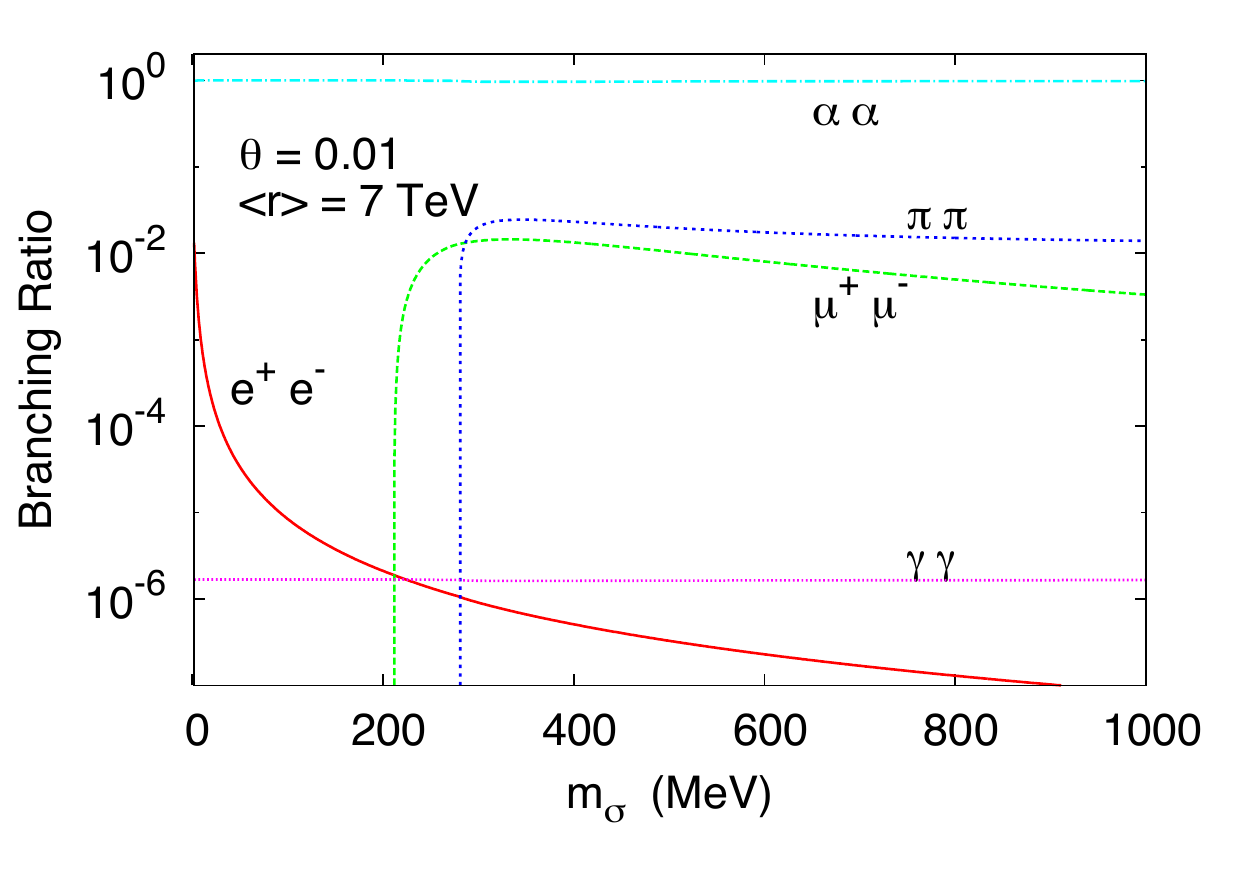}
\caption{\small \label{decay}
Decay branching ratios for the $\sigma$ field for (a) $\langle r \rangle =
1$ TeV and (b) $\langle r \rangle = 7$ TeV. The mode $\pi \pi$ includes
$\pi^+\pi^-$ and $\pi^0\pi^0$. The mixing parameter $\theta$ is set at  0.01.
}
\end{figure}

The major difference is the decay mode $\sigma \to \alpha \alpha$
of the Goldstone boson.
The partial width is
\begin{equation}
\Gamma (\sigma \to \alpha \alpha ) = \frac{ m_\sigma^3}{ 32 \pi \langle r
 \rangle^2 } \;.
\end{equation} 
It is easy to see that the visible partial widths are all proportional
to $\theta^2$ because the decay into visible particles is only possible 
via the mixing with the SM Higgs boson. On the other hand, the decay
into a pair of Goldstone bosons is not suppressed by the 
mixing angle, but inversely 
proportional to the square of $\langle r \rangle$. We show the branching
ratios of the $\sigma$ field for $\langle r \rangle$ = 1 and 7 TeV in 
Fig.~\ref{decay}.

Mostly, the $\sigma$ field decays invisibly into the Goldstone bosons.
It essentially adds to the invisible width of the SM Higgs boson, as the
Higgs boson can decay via $H \to \sigma \sigma \to 4 \alpha$ and
$H \to \alpha \alpha$. 
However, when $\langle r \rangle$ goes to a
very large value, say 7 TeV, as remarked already in \cite{weinberg}, 
the decay of $\sigma \to \pi \pi$ can be as large as 2\%.

We show the ratio
\begin{equation}
\label{f}
f \equiv \frac{\Gamma(\sigma \to \pi \pi)}{\Gamma(\sigma \to \alpha\alpha)}
=  \theta^2 \frac{4}{27} \frac{\langle r \rangle^2}{\langle \phi \rangle^2} \,
  \left( 1- \frac{4 m_\pi^2}{m_\sigma^2} \right )^{1/2} \,
  \left( 1 +  \frac{11 m_\pi^2}{2 m_\sigma^2} \right )^{2}
\end{equation}
in Fig.~\ref{limit} for $m_\sigma =$ 500 MeV. 
In most part of the allowed region, the ratio $f$ is
well below $10^{-4}$, thus mostly the $\sigma$ field decays into Goldstone
boson. Nevertheless, if one goes to the corner where 
$\frac{\langle \phi \rangle}{\langle r \rangle}$ is very small, we can 
achieve $f \approx 10^{-2}$.  Such a value of $f$ would imply
very interesting signatures for the $\sigma$ field and the Goldstone 
boson.

\section{Collider Signatures}

When the branching ratio $B(\sigma \to \pi \pi) \approx 2\%$, the
collider signature would be very interesting. The dominant production
of the $\sigma$ field is via the decay of the Higgs boson, followed by
the decays of the two $\sigma$ fields. We can look for one $\sigma$
decaying invisibly into a pair of Goldstone bosons while the other one
decays visibly into a pair of pions.  Therefore, we expect
\begin{equation} 
 gg \to H \to \sigma \sigma \to (\pi \pi) (\alpha \alpha) \;,
\end{equation}
where the invariant mass of the pion pair is located right at $m_\sigma$.  The
signature would be a distinguished pion pair with $m_{\pi\pi} \approx
m_\sigma$ plus a large missing energy carried away by the Goldstone
bosons $\alpha$.

We perform a rough estimate of event rate here. The production cross section
of the SM Higgs boson the LHC-8 is about 19 pb \cite{wikicern}, 
and the non-standard
decay branching ratio of the Higgs boson 
is limited to be less than about 20\% \cite{higgcision}.
Therefore, using the analysis above we choose a currently allowed 
branching ratio of the Higgs boson:
\begin{equation}
B(H \to \sigma\sigma) \alt 10\% \;.
\end{equation}
The cross section at the LHC-8 with $\langle r \rangle = 7$ TeV 
would be
\begin{eqnarray}
&& \sigma(gg\to H)\times B(H \to \sigma\sigma) \times B(\sigma \to \pi\pi)
\times B(\sigma \to \alpha\alpha) \times 2 \nonumber \\
& & \approx  19\;{\rm pb} \times 0.1 \times 0.02 \times 0.97 \times 2
 \approx 73 \; {\rm fb}\;.
\end{eqnarray}
For LHC-14, one should multiply the above number by a factor of 
$2.8$. 

Since the intermediate $\sigma$ boson is only $O(1)$ GeV, its decay
products would be very collimated. The two $\alpha$'s become missing
energies, while the two pions are very collimated, which 
appear to be a ``microjet'', and experimentally 
it looks like a $\tau$ jet. The final state then consists of a microjet jet, 
which is made up of two pions, and a large missing energy. 
We first discuss the case when the two pions are charged pions.
Ideally,  we would like to separate the two charged pions with
an angular separation between them of order
$\sim 2 m_\sigma / p_{T_\sigma} = 1\, {\rm GeV} / 60\,{\rm GeV}
\approx 0.015$ which is rather small.
 Only the pixel detector inside the LHC experiments
has some chances of separating them.
The pixel tracker of the CMS detector~\cite{CMS}
consists of three barrel layers with radii 4.4, 7.3 and 10.2 cm,
and two endcap disks on each side of the barrel section. 
The spatial resolution ranges from 20 $\mu$m to about 100 $\mu$m, depending
on the direction. 
Taking conservatively 100 $\mu$m as the spatial resolution
and divide it by the average radius of the pixel detector, say 5 cm, 
we obtain an angular resolution of  $2\times 10^{-3}$
\footnote
{ With both outer trackers and pixel detectors, the resolution
could be $2-10$ times better than $2\times 10^{-3}$ for pions with
$p_T  > 10$ GeV.
}.
This is smaller than the average angular separation 
between the two charged pions estimated above by almost an order of magnitude.
Thus it seems quite plausible to separate the two collimated charged pions.
However, there is no guarantee that the pattern recognition algorithms would be able to
reconstruct two distinct tracks, especially in the presence of large number of 
pile-up events.
In the next phase of the CMS, another layer will be added to the pixel detectors
at a radius of 16 cm~\cite{CMS-2}. The angular resolution will
be further improved and the likelihood of separating the tracks of the two charged pions 
will be increased.

If the experiment cannot resolve the two charged pions, 
then the final state 
will look like a single jet consisting of some hadrons, 
plus missing energy. It is similar to signatures of many new models beyond the SM. 
In this case, one can make use of the associated production of the Higgs boson 
with a $W$ (or $Z$) boson, followed by the leptonic decay of the $W$
and the same decay mode of the Higgs boson:
\begin{equation}
   pp \to  W H \to (\ell \nu) (\sigma \sigma) \longrightarrow 
      (\ell \nu) (\pi \pi + \alpha \alpha)  \;.
\end{equation}
The final state then consists of a charged lepton, a single jet of two unresolved
charged pions, plus missing energy.  
The charged lepton is an efficient trigger of the events.
The major SM background is the production of $W+1$ jet, which could be orders
of magnitude larger \cite{w1j}.  It presents an extreme challenge
for experimentalists, although we may make use of the missing 
energy spectrum, because the signal also receives missing energy from
$\sigma \to \alpha \alpha$ decay in addition to the neutrino from the $W$ decay. 
One may also use the feature of a microjet (similar to a $\tau$ jet) that is somewhat ``thin'' compared to
the usual hadronic jet to separate the signal from backgrounds. 
In the case that one of the $\sigma$'s decays into two neutral pions, 
the process can give rise to 4 photons collimated as one ``fat'' photon.
The final state would be a charged lepton, a ``fat'' photon plus missing energy,
challenged by the major SM background of $W \gamma$ production which has a much larger 
cross section \cite{wg}. However, one can make use of the fact
that the photon in the signal is ``fat'' to distinguish it from the 
background one.

Therefore, using both the gluon fusion and associated production with a $W$
we have provided more options to explore this model.
However, in all situations that we studied above, they present great challenges to the experimentalists.
Detailed detection simulations are needed in order to settle down
if the proposed search is feasible or not. 
In the following, we will be contended by performing rough estimates 
on the signal cross section 
at the LHC-8 and LHC-14, so that experimentalists can have some ideas
how large the signal cross section that one can obtain.

At $\langle r \rangle =7$~TeV, 
the branching ratio of $\sigma$ into $\pi\pi$ is as large as 2\%.
For smaller $\langle r \rangle=3 - 7$ TeV,
the branching ratio into $\pi\pi$ ranges from about 0.4\% to 2\%,
for which  we may have enough cross sections for detection. 
We perform parton-level Monte Carlo simulations to 
estimate the event cross sections 
at the LHC-8 and LHC-14 for $\langle r \rangle = 3 - 7$ TeV.
We normalize the uncut gluon fusion cross sections and the 
associated production cross sections to those
given in the LHC Physics Web site \cite{wikicern}.
For both the pions and charged lepton, 
we impose the same $p_T$ and rapidity cuts  
as $p_T > 30 \;{\rm GeV}$ and $|\eta| < 2.5$ respectively.
We show the cross sections after the cuts in Table~\ref{x-sec}.
We have multiplied the cross sections by the branching ratios
$B(H \to \sigma \sigma) \times B(\sigma \to \pi \pi) \times
B(\sigma \to \alpha \alpha) \times 2$ to the Higgs boson
decay, and $B(W\to \ell \nu) = 2/9$ to the $W$ boson decay.
At the LHC-8 with about 20 fb$^{-1}$, the gluon fusion can produce 
a handful of events against the background if the two pions can be resolved.
Nevertheless, if the pions cannot be resolved the associated production
only has a cross section of order $O(0.05)$ fb, which may not be
enough for detection.  At the LHC-14 with a projected luminosity of 
$O(100)$ fb$^{-1}$, both the gluon fusion and associated production
give sizable event rates whether or not the two pions can be resolved.
Here, as mentioned above, the most important experimental issue is 
resolving the two pions. Although our rough estimate of angular separation
by the pixel and tracking detectors indicates
 one may be able to resolve the pions, 
difficulties coming from the pile-up, pattern recognition, and track reconstruction
post real challenges for our experimentalists.  
A proper detector simulation is called for before any realistic conclusion can be drawn.

\begin{table}[tbh!]
\caption{\small \label{x-sec}
Cross sections in fb for the gluon fusion process 
$pp \to H \to \sigma\sigma \to (\pi\pi) (\alpha\alpha)$
and the associated process
$pp \to WH \to (\ell \nu) (\sigma\sigma) \to (\ell \nu) (\pi\pi + \alpha\alpha)$
at the LHC-8 and LHC-14 with the selection cuts described in the text.
We choose $m_\sigma = 500$ MeV.
}
\centering
\begin{ruledtabular}
\begin{tabular}{cccccc}
$\langle r \rangle$ & $B(\sigma \to \pi \pi)$ & 
       \multicolumn{2}{c} {Cross Section (fb) LHC-8} &
       \multicolumn{2}{c} {Cross Section (fb) LHC-14} \\
(TeV) &   &  gluon fusion & $WH$ &  gluon fusion & $WH$ \\
\hline
3   & $3.72\times 10^{-3}$ & 0.16  & 0.013  &  0.39  & 0.024 \\
4   & $6.58\times 10^{-3}$ & 0.27  & 0.022  & 0.68 & 0.043 \\
5   & $1.02\times 10^{-2}$ & 0.42   & 0.034  & 1.05  & 0.067 \\
6   & $1.46\times 10^{-2}$ & 0.60  & 0.049  & 1.50  & 0.095 \\
7   & $1.97\times 10^{-2}$ & 0.80  & 0.065  &  2.00  & 0.13   
\end{tabular}
\end{ruledtabular}
\end{table}

To summarize, the logical possibility of the existence of a hidden sector of 
Goldstone bosons masquerading as fractional cosmic neutrinos and communicate
to our visible world through the Higgs portal as
suggested recently by Weinberg
\cite{weinberg} is explored further phenomenologically here.  We have
studied the constraints from the invisible Higgs search at LEP-II, the
invisible Higgs width derived from global fittings using all the LHC
signal strength data, and the condition of muon decoupling from
evolution of our Universe.  We also studied Higgs decays into a pair
of $\sigma$ and its various decay modes.  This interesting idea of
Goldstone bosons as cosmic neutrino impostors can be tested by
searching for the process of $gg \to H \to \sigma \sigma \to (\pi \pi)
(\alpha \alpha)$ and the associated production
$WH \to (\ell \nu) (\sigma \sigma) \to (\ell\nu) (\pi \pi + \alpha \alpha)$
at the LHC-8 and LHC-14.

\newpage

\section*{Acknowledgments}
We thank Kevin Burkett, Kai-Feng Chen, Shih-Chieh Hsu, and Shin-Shan Yu
for useful discussions regarding experimental issues, and also
thank Chih-Ting Lu for pointing out the correct formula for $\sigma$ 
decays into pion pair.
This work was supported in parts by the National Science Council of
Taiwan under Grant Nos. 99-2112-M-007-005-MY3, 102-2112-M-007-015-MY3, and
101-2112-M-001-005-MY3, in part by U.S. Department of Energy under 
DE-FG02-12ER41811, as well as the
WCU program through the KOSEF funded by the MEST (R31-2008-000-10057-0). 
WYK would like to thank the hospitalities of Institute of Physics, 
Academia Sinica and the Physics Division of NCTS.


\begin{thebibliography}{99}
\bibitem{wimpmode} In \cite{weinberg}, Weinberg also suggested the
  global $U(1)$ symmetry is the WIMP number by adding in the model one
  single Dirac field $\psi$ with $W=+1$ while the $S$ field has $W=+2$
  so that its VEV leaves an unbroken $Z_2$ symmetry $\psi \to
  -\psi$. The physical spectrum will contain two Majorana fields
  $w_{\pm}$.  The one with lighter mass $w_-$ could then be the dark
  matter candidate which has an interaction
  $-\frac{f\theta}{2}\overline {w_-} w_- H$ with the Higgs boson
  through mixing.  If $2 m_{w_-} < m_{H}$, it will contribute to the
  invisible Higgs decay width as $\Gamma_{H \to \overline {w_-} w_-} =
  (1/16\pi)(f g \langle \phi \rangle \langle r \rangle / (m^2_H -
  m^2_\sigma) )^2 ( m^2_H - 4 m^2_{w_-} )^{3/2}$. The exponent 3/2
  reflects the P-wave feature of the decay amplitude.  We do not
  include this in our analysis since more free parameters are
  necessary to bring in.
\end{thebibliography}
\end{document}